\documentclass[useAMS,usenatbib]{mnras}

\usepackage{amsmath}
\usepackage{graphicx}
\usepackage{longtable}
\usepackage{amssymb}
\usepackage{tabularx, blindtext}
\usepackage{lscape}
\usepackage{url}
\newcommand{\feh}{\ensuremath{\rm [Fe/H]}}
\newcommand{\teff}{\ensuremath{T_{\rm eff}}}
\newcommand{\logg}{\ensuremath{\log{g}}}
\newcommand{\zaspe}{\texttt{ZASPE}}
\newcommand{\vsini}{\ensuremath{v \sin{i}}}
\newcommand{\kms}{\ensuremath{\rm km\,s^{-1}}}
\def\MgI{\ion{Mg}{I}}

\title[Zonal Atmospheric Stellar Parameters Estimator]{\zaspe: A Code to Measure Stellar Atmospheric Parameters and their Covariance from Spectra}
\author[Brahm et~al.]{Rafael Brahm$^{1,2}$\thanks{E-mail:
rbrahm@astro.puc.cl, The code can be found in the following url: \url{http://github.com/rabrahm/zaspe}}, Andr\'es Jord\'an$^{1,2}$, Joel Hartman$^3$, Gaspar Bakos$^{3,4,5}$\\
$^{1}$Instituto de Astrof\'isica, Facultad de F\'isica, Pontificia Universidad
Cat\'olica de Chile, \\
Av. Vicu\~na Mackenna 4860, 7820436
Macul, Santiago, Chile.\\
$^{2}$Millennium Institute of Astrophysics, Santiago, Chile.\\
$^3$Departmen of Astrophysical Sciences, Princeton University, NJ 08544, USA.\\
$^4$Alfred P.~Sloan Research Fellow\\
$^5$Packard Fellow
}
\begin{document}

\date{Draft Version 4.0}

\pagerange{\pageref{firstpage}--\pageref{lastpage}} \pubyear{2002}

\maketitle

\label{firstpage}

\begin{abstract}
We describe the Zonal Atmospheric Stellar Parameters Estimator (\zaspe), a new algorithm, and its associated code, for determining precise stellar atmospheric parameters and their uncertainties from high resolution echelle spectra of FGK-type stars. \zaspe\ estimates stellar atmospheric parameters by comparing the observed spectrum against a grid of synthetic spectra only in the most sensitive spectral zones to changes in the atmospheric parameters. Realistic uncertainties in the parameters are computed from the data itself, by taking into account the systematic mismatches between the observed spectrum and the best-fit synthetic one. The covariances between the parameters are also estimated in the process. \zaspe\ can in principle use any pre-calculated grid of synthetic spectra. We tested the performance of two existing libraries \citep{coelho:2005, husser:2013} and we concluded that neither is suitable for computing precise atmospheric parameters. We describe a process to synthesise a new library of synthetic spectra that was found to generate consistent results when compared with parameters obtained with different methods (interferometry, asteroseismology, equivalent widths).
\end{abstract}


\section{Introduction}
\label{sec:intro}

The determination of the physical parameters of stars is a fundamental requirement
for studying their formation, structure and evolution.
Additionally, the physical properties of extrasolar planets depend strongly on how well
we have characterised their host stars.
In the case of transiting planets, the measured transit depth is related to the ratio of the planet to stellar radii.
Similarly, for radial velocity planets the semi-amplitude of the orbit is a function of
both the mass of the star and the mass of the planet.
In the case of directly imaged exoplanets, their estimated masses depend on the age
of the systems.
With more than 3000 planets and planetary candidates discovered, mostly by the
\textit{Kepler} mission \citep[e.g.][]{howard:2012, burke:2014}, homogeneous and
accurate determination of the physical parameters of the host stars are required for
linking their occurrence rates and properties with different theoretical predictions
\citep[e.g.][]{howard:2010, buchhave:2014}.

Direct determinations of the physical properties of single stars (mass, radius and age)
are limited to a couple dozens of systems.
Long baseline optical interferometry has been used on bright sources with known
distances to measure their physical radii \citep{boyajian:2012,boyajian:2013} and
precise stellar densities have been obtained using asteroseismology on stars
observed by \textit{Kepler} and \textit{CoRoT} \citep[e.g.][]{silva:2015}.
Unfortunately, for the rest of the stars, including the vast majority of planetary hosts,
physical parameters can not be measured and indirect procedures have to be
adopted in which the atmospheric parameters, such as the effective temperature
(\teff), surface gravity (\logg) and metallicity (\feh), are derived from stellar
spectra by using theoretical model atmospheres. Stellar evolutionary models are then
compared with the estimated  atmospheric parameters in order to determine the
physical parameters of the star.

The amount of information about the properties of the stellar atmosphere contained in
its spectrum is enormous.
Current state-of-the-art high resolution echelle spectrographs are capable
of detecting subtle variations of spectral lines which, in principle, can be translated into the determination of 
the physical atmospheric conditions of a star with exquisite precision. However,
there are several factors that reduce the precision that can be achieved.
On one side, there are many other properties of a star that can
produce changes on the absorption lines. For example, velocity fields on the surface of the star,
which include the stellar rotation (which may be differential) and the micro and macro
turbulence, modify the shape of the absorption lines. Non-solar abundances change the
particular strength of the lines of each element. Therefore, in order to obtain
precise atmospheric parameters, all of these variables have to be considered.
However, even when all the significant variables of the problem are taken into account, the precision in the
parameters becomes limited by modelling uncertainties, e.g.: imperfect modelling of the stellar atmospheres and spectral
features due to unknown opacity distribution functions, uncertainties in the properties of particular atomic and molecular
transitions, effects arising from the assumed geometry of the modelled atmosphere and non-LTE effects.
These sources of error are unavoidable and are currently the main problem for obtaining reliable
{\em uncertainties} in the estimated stellar parameters. Most of the actual algorithms
that compute atmospheric parameters from high resolution stellar spectra do not consider in detail
this factor for obtaining the uncertainties. The problem is that if the reported uncertainties are unreliable,
then they propagate to the planetary parameters and can bias the results or hide potential trends in the properties of
the system under study that, if detectable, could lead to deeper insights into their formation and evolution.

A widely used procedure for obtaining the atmospheric parameters of a star 
consists in comparing the observed spectra against synthetic
models and adopt the parameters of the model that produces the best match as the 
estimated atmospheric parameters of the observed star.  This technique has been
implemented in algorithms  such as \texttt{SME} \citep{valenti:96}, \texttt{SPC} \citep{buchhave:2012}, and \texttt{iSpec} \citep{blanco:2014} to derive parameters 
of planetary host stars. Thanks to the large number of spectral features used, this method has been shown
to be capable of dealing with spectra having low SNR, moderate resolution and a wide range
of stellar atmospheric properties. However, one of the major drawbacks of spectral synthesis
methods for the estimation of the atmospheric parameters is the determination of their uncertainties.
This problem arises because the source of error is not the Poisson noise of the observed spectrum, but
instead is usually dominated by imperfections in the synthesised model spectra, which produce highly
correlated residuals. In such cases standard procedures for computing uncertainties for the parameters are not reliable.
For example, \texttt{SPC} computes the internal uncertainties using the dispersion from different measurements
in the low SNR regime, but an arbitrary floor is applied when the uncertainties are expected to be dominated by the
systematic miss-matches between models and data.
Additionally, \cite{torres:2012} showed that there are strong correlations between the atmospheric parameters
obtained using spectral synthesis techniques, and therefore, the covariance matrix of the parameters should be
a required output of any stellar parameter classification tool so that the uncertainty of its results are properly propagated to the posterior inferences that are made using them. Recently, \citet{czekala:2014} introduced \texttt{Starfish}, a code that allows robust estimation of stellar parameters using synthetic models by using a likelihood function with a covariance structure described by Gaussian processes. \texttt{Starfish} allows robustness to synthetic model imperfections through a principled approach using a sophisticated likelihood function and provides full posterior distributions for the parameters, but as we will argue later its uncertainties are significantly underestimated.

In this paper we present a new algorithm, dubbed Zonal Atmospheric Stellar Parameters Estimator (hereon \zaspe) for estimating stellar atmospheric parameters using the
spectral synthesis technique. The uncertainties and correlations in the parameters are computed from the data itself
and include the systematic mismatches due to the imperfect nature of the theoretical spectra. The structure of the paper is as follows. In \S~\ref{method}
we describe the method that \zaspe\ uses for determining the stellar parameters and their covariance matrix, including details on the synthesis of a new synthetic library to overcome limitations of the existing libraries for stellar parameter estimation.
In \S~\ref{ssec:results} we summarise the performance of \zaspe\ on a sample of stars with measured stellar parameters, and we compare our uncertainties with those produced by \texttt{Starfish}. Finally, in \S~\ref{sec:sum} we summarize and conclude.

\begin{table*}
\label{grid}
 \centering
 \begin{minipage}{180mm}
  \caption{Grid extension and spacing for each ZASPE iteration.}
  \begin{tabular}{@{}cccccccccc@{}}
  \hline
   
   Iteration     &   $\teff^{i}$ [K]  & $\teff^f$  [K] &  $\Delta \teff$  [K] &  $\logg^{i}$ &  $\logg^f$ & $\Delta\logg$ &  $\feh^i$ & $\feh^f$ &  $\Delta\feh$\\
 \hline
 1 & 4000 & 7000 & 200 & 1.0 & 5.0 & 0.5 & -1.0 & 0.5 & 0.5 \\
 2 & $\teff^{c}$ - 500 & $\teff^{c}$ + 500 & 100 & $\logg^{c}$ - 0.6 & $\logg^{c}$ + 0.6 & 0.2 & $\feh^c$ - 0.4 & $\feh^c$ + 0.4& 0.1 \\
 3 & $\teff^{c}$ - 300 & $\teff^{c}$ + 300 & 75  & $\logg^{c}$ - 0.6 & $\logg^{c}$ + 0.6 & 0.2 & $\feh^c$ - 0.3 & $\feh^c$ + 0.3& 0.075 \\
 4 & $\teff^{c}$ - 200 & $\teff^{c}$ + 200 & 50  & $\logg^{c}$ - 0.4 & $\logg^{c}$ + 0.4 & 0.1 & $\feh^c$ - 0.2 & $\feh^c$ + 0.2& 0.05 \\
$>$ 4 & $\teff^{c}$ - 50 & $\teff^{c}$ + 50    & 10   & $\logg^{c}$ - 0.2 & $\logg^{c}$ + 0.2 & 0.05 & $\feh^c$ - 0.06 & $\feh^c$ + 0.06& 0.02 \\
\hline
\end{tabular}
\end{minipage}
\end{table*}

\section[]{The Method}
\label{method}
In order to determine the atmospheric stellar parameters of a star,
\zaspe\ compares an observed {\em continuum normalised} spectrum
against synthetic spectra using least squares minimisation by performing an
iterative algorithm that explores the complete parameter space of FGK-type stars.
For simplicity, we assume first that we are able to generate an unbiased
synthetic spectrum with any set of stellar atmospheric parameters
(\teff, \logg\ and \feh). By unbiased we mean that there are
no systematic trends in the level of mismatch of the synthesised and real spectra as
a function of the stellar parameters, but there can be systematic
mismatches that are not a function of stellar parameters. If $F_{\lambda} $
is the observed spectrum and $S_{\lambda} (\vec{\theta})$ is
the synthesised spectrum with parameters $\vec{\theta} = \{\teff, \logg, \feh\}$,
the quantity that we minimize is
\begin{equation}
\label{dif}
   X^2(\vec{\theta}) = \sum_{\lambda} ( F_{\lambda} - S_{\lambda} (\vec{\theta}) )^2.
\end{equation}
in Equation~\ref{dif} we have not included the weights coming from the uncertainties in the observed flux 
because we are assuming that the signal to noise ratio (SNR) of the data is
high enough for the uncertainties in the parameters to be governed by the systematic mismatches
between the data and the models.

The synthesised spectrum needs to have some processing done in order to compare  
it against the observed one. 
We do not treat microturbulence and macroturbulence as free parameters,
but instead we assume that these values are functions of the atmospheric
parameters. The microturbulence value is required during the
process of synthesising the spectra and it depends on the particular 
spectral library selected to do the comparison (see \S~\ref{grids}).
On the other hand, the macroturbulence degradation is applied after the
synthetic spectra have been generated. We compute the macroturbulence
value for each synthetic spectrum from its \teff\ using the empirical relation
given in \cite{valenti:2005}\footnote{As was pointed out by \cite{torres:2012},
the formula in \cite{valenti:2005} has a wrong sign.}, namely:
\begin{equation}
v_{\rm mac} = \left( 3.98 + \frac{T_{eff}-5770 K}{650 K} \right) \textnormal{km s}^{-1}. 
\end{equation}

The effect of macroturbulence on the spectrum is given by a convolution
 with a Gaussian kernel whose standard deviation is given by $\sigma_{mac}=0.297 v_{mac}$,
 as was approximated in \cite{takeda:2008}.
The degradation to the particular instrumental resolution, $R = \Delta \lambda / \lambda$ is performed
by convolving the synthetic spectrum with another Gaussian kernel whose standard
deviations is $\sigma_{res}=\lambda/(2.3R)$.
The model spectrum is then split according to the echelle orders of the
observed spectrum and the pixelization effect is taken into account by integrating
the synthetic flux over each wavelength element of the observed spectrum.

\subsection{The sensitive zones}
\label{sec:zones}
One of the novel features of \zaspe\ in contrast to other similar codes,
is that the comparison between the observed and synthetic spectra is
performed in particular optimized wavelength zones, rather than using the full spectrum.
These zones correspond to the most sensitive regions of the spectra
to changes in the stellar parameters and are redefined in each iteration of \zaspe.

\begin{figure}
 \includegraphics[width=\columnwidth]{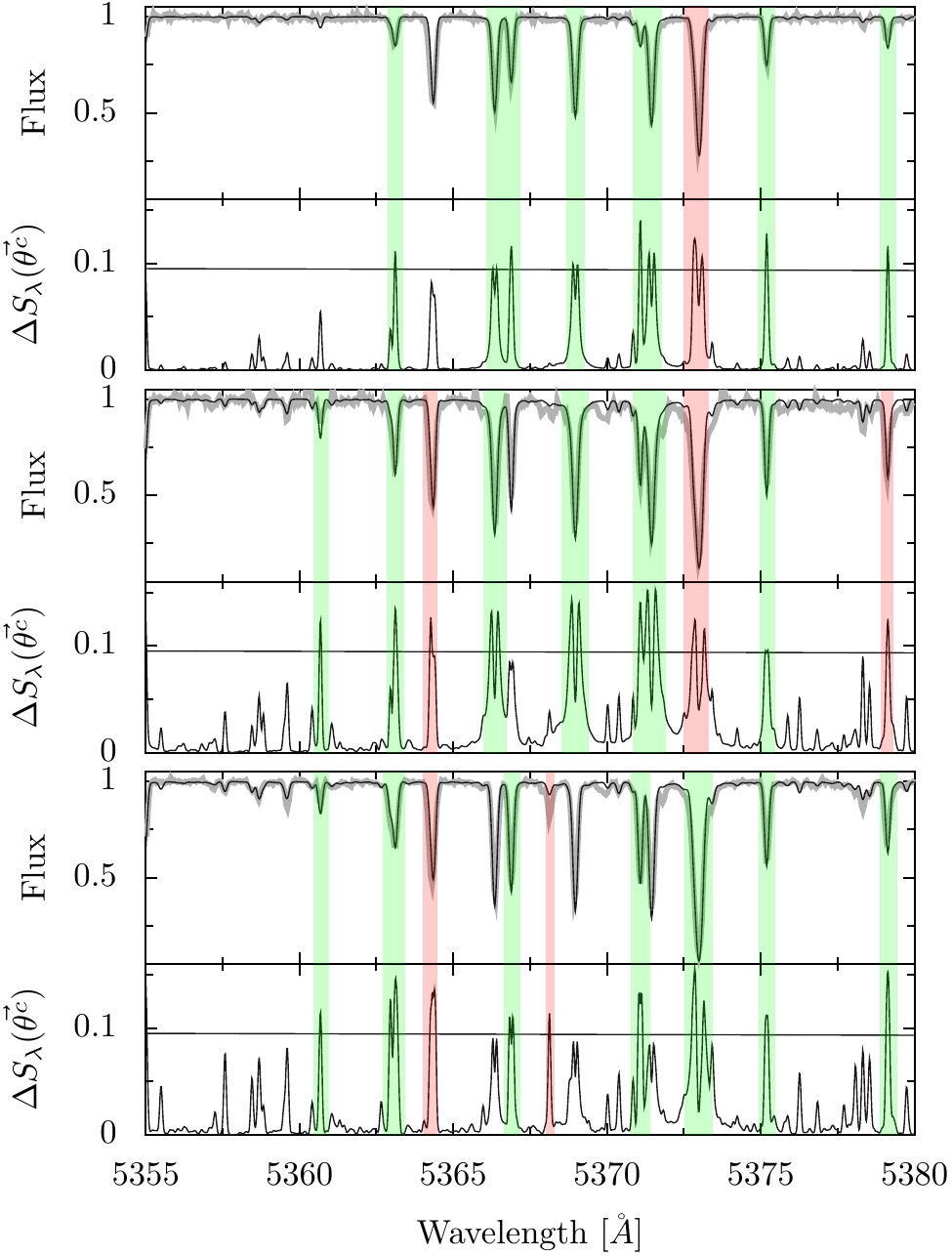}
 \caption{Sensitive zones determined by \zaspe\ in a portion of the wavelength coverage for three different stars (top panel: late F-dwarf, central panel: solar-type star, bottom panel: K-giant). In each panel the superior plot corresponds to the observed spectrum (thick line) and the optimal synthetic one determined by ZASPE (thin line), while the inferior plot shows the gradient $\Delta S_{\lambda}(\vec{\theta^c})$ of the synthetic grid evaluated at the parameters of the optimal synthetic spectrum, and the threshold (horizontal line) that determines which regions of the spectrum are defined as sensitive zones. The green coloured regions correspond to the sensitive zones determined by ZASPE where the comparison between data and models is performed. The red coloured regions are regions of the spectrum that are initially identified as sensitive zones by ZASPE but then rejected because the average residual between the optimal model and the data in these particular regions is significantly higher (greater that $3\sigma$) than in the rest of the sensitive zones.}
 \label{zones}
\end{figure}

These sensitive regions are determined from the approximate gradient of the
modelled spectra with respect to the stellar parameters at $\vec{\theta^c}$, 
where $\vec{\theta^c} = \{\teff^c, \logg^c, \feh^c\}$ is the set of 
parameters that produced the minimum $X^2$ in the previous iteration.
In practice, once $\vec{\theta^c}$ is determined, \zaspe\ computes the following
finite differences:

\begin{align}
   \Delta S_{\teff}^1 = \|S_\lambda(\teff^c+200,\logg^c,\feh^c) - S_\lambda(\vec{\theta^c}) \|, \\
   \Delta S_{\teff}^2 = \|S_\lambda(\teff^c-200,\logg^c,\feh^c) - S_\lambda(\vec{\theta^c}) \|, \\
   \Delta S_{\logg}^1 = \|S_\lambda(\teff^c,\logg^c+0.3,\feh^c) - S_\lambda(\vec{\theta^c}) \|, \\
   \Delta S_{\logg}^2 = \|S_\lambda(\teff^c,\logg^c-0.3,\feh^c) - S_\lambda(\vec{\theta^c}) \|, \\
   \Delta S_{\feh}^1 = \|S_\lambda(\teff^c,\logg^c,\feh^c+0.2) - S_\lambda(\vec{\theta^c}) \|, \\
   \Delta S_{\feh}^2 = \|S_\lambda(\teff^c,\logg^c,\feh^c-0.2) - S_\lambda(\vec{\theta^c}) \|,
\end{align}

\noindent from which the approximate gradient of the synthesised spectra with respect to the
atmospheric parameters, averaged on the three parameters, is estimated as
\begin{multline}
\Delta S_{\lambda}(\vec{\theta^c}) = \frac{1}{6} (\Delta S_{\teff}^1 + \Delta S_{\teff}^2 + \Delta S_{\logg}^1 \\ 
+ \Delta S_{\logg}^2 + \Delta S_{\feh}^1 + \Delta S_{\feh}^2).
\end{multline}
Spectral regions where $\Delta S_{\lambda}(\vec{\theta^c})$ is greater than an predefined threshold are identified as
the sensitive zones, which we denote as $\{z_i\}$. Figure~\ref{zones} shows a portion of the spectrum for three different
stars and the sensitive zones selected in the final \zaspe\ iteration in each case. It can be seen that the selected sensitive
zones correspond to the spectral regions where absorption  lines are present, but that not all the absorption lines are identified
as sensitive zones at a given threshold. In addition, the regions that are selected as sensitive zones vary according to the
properties of the observed star.
For identifying the sensitive zones we have introduced four quantities that take arbitrary values. These correspond to the
distance in  step sizes for the three atmospherical parameters (200 K, 0.3 dex and 0.2 dex for \teff, \logg\ and \feh, respectivelyy)
that are used to compute the gradient of the grid, and the threshold value (0.09 as default) that defines as sensitive zones the
spectral regions where the gradient is greater than this value.
The particular values that we selected as default allow \zaspe\ to identify a great number of sensitive zones even for F-type stars,
but at the same time each of these zones counts only with one or two significant absorption lines even in the case of the crowded K-type stars.
This last requirement is mandatory for the procedure that \zaspe\ uses to compute the  errors in the parameters (see \S~\ref{errors}).
At the same we chose step sizes in the parameters that are slightly larger than the expected errors but small enough to avoid the emergence
other significant features in the spectra of each sensitive zone.


The introduction of the zones into the problem allows also the rejection of portions
of the spectra that strongly deviate from $S_{\lambda} (\vec{\theta^c})$, due to
modelling problems or by the presence of artifacts in the data that remain in the spectrum(e.g., cosmic rays, bad columns).
In practise, outliers are identified by computing the root mean square (RMS) of the residuals between the
observed spectra and the optimal synthetic one in each sensitive zone and zones with RMS values
greater than 3 times the average RMS value are rejected.
Once the sensitive zones are known, \zaspe\ builds a binary mask, $M_{\lambda}$, 
filled with ones in the spectral range of the sensitive zones and zeros elsewhere, i.e.
\begin{equation}
M_\lambda = \left\{ 
  \begin{array}{lr}
       1  : \lambda \in \{z_n\}\\
       0  : \lambda \not \in \{z_n\}
   \end{array} \right.
\end{equation}
For the next iteration, the function to be minimised will be
\begin{equation}
\label{chism}
   X^2(\vec{\theta}) = \sum_{\lambda} M_{\lambda} ( F_{\lambda} - S_{\lambda} (\vec{\theta}) )^2.
\end{equation}
In the first iteration of \zaspe\ the complete spectral range is utilised and $M_\lambda \equiv 1\,\,\, \forall \lambda$.

\subsection{Continuum normalization}
\begin{figure}
 \includegraphics[width=\columnwidth]{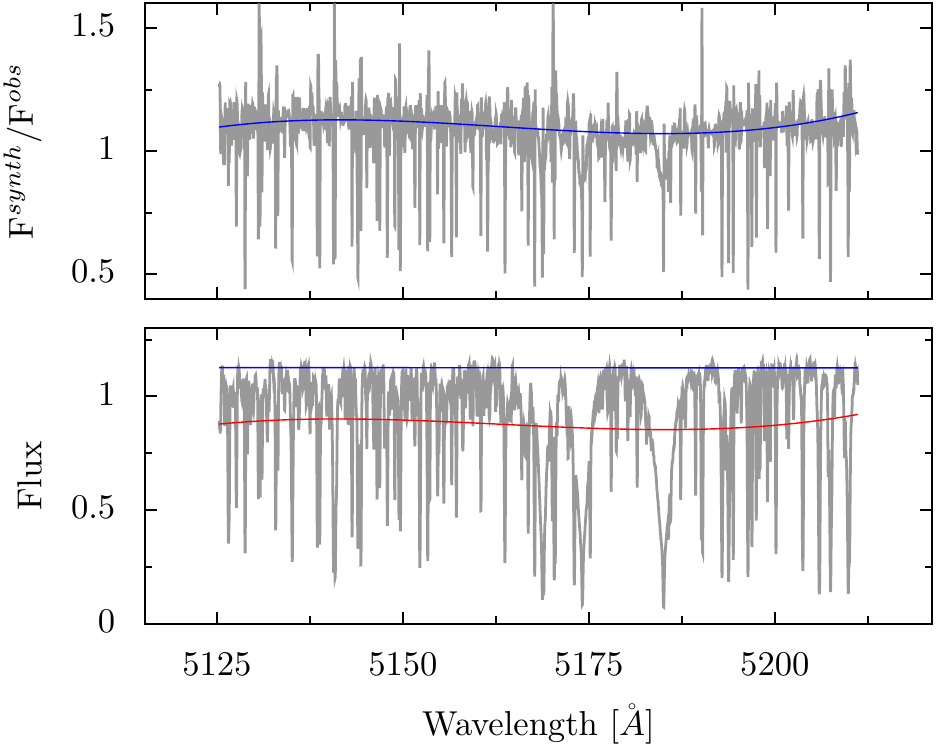}
 \caption{Top: The blue line corresponds to the polynomial fitted to the ratio between the optimal synthetic spectrum find in the previous \zaspe\ iteration and the observed spectrum. This procedure allows to determine the continuum normalisation without overfitting wide spectral features. Bottom: comparison between the continuum determined by the algorithm that \zaspe\ uses (blue line) and the one determined by fitting a simple polynomial (red line), which clearly is heavily affected by the presence of strong absorption features.}
 \label{cont}
\end{figure}

 \zaspe\ contains an algorithm that performs the continuum normalisation of the observed spectra,
which is required for a proper comparison with the synthetic spectral library.
One important assumption that we make at this step is that the large scale variations of the
observed flux as function of wavelength must be smooth and it must be possible
to accurately trace them with a simple low degree polynomial. This means that
the observed spectrum should be at least corrected by the blaze function and it
should not contain systematics in order to define a proper continuum or
pseudeo-continuum.
If the input observed spectrum satisfies this constrain, then our continuum normalisation
algorithm deals with the presence of both, shallow and strong spectral features.
The continuum is updated after each \zaspe\ iteration, because the optimal model
is used by the algorithm to avoid an overfiting of the wide spectral features, like the
zone of the \ion{Mg}{I}b triplet for example. The idea is to bring the continuum of the
observed spectrum to match the continuum of the optimal synthetic one. Therefore, for every echelle order,
the optimal synthetic spectrum found after each \zaspe\ iteration is divided by the observed
spectrum, and polynomials are fitted to these ratios using an iterative procedure that 
rejects regions where the model and data significantly differ. Given that both model
and data should contain the wide spectral features, these disappear when
the division is performed, and the only significant features that remain 
are the instrumental response and the black body wavelength dependence of the observed spectrum.
The polynomials obtained  for each echelle order are then multiplied by the
observed spectrum, which corrects for the large scale smooth variations.
Finally, a straight line is fitted to this corrected spectrum using an iterative process 
that excludes the absorption lines from the fit. This last  normalisation
is applied to ensure that the continuum or pseudo-continuum takes values equal to 1,
which is particularly important when determining and applying the mismatch factors
of \S~\ref{errors}. Additionally, the synthetic spectra are also normalised by a straight line.
Figure~\ref{cont} shows that the normalisation algorithm used by \zaspe\ performs
better in zones with wide spectral features than a simple polynomial fit.

\subsection{Radial velocity and \vsini}

In each \zaspe\ iteration, the search of the $X^2$ minimum is performed simultaneously over the three atmospheric parameters.
However, the velocity of the observed spectrum with respect to the synthesised spectra (radial velocity) and the \vsini\
value are updated in each \zaspe\ iteration after $\vec{\theta^c}$ is determined, because of the slight dependence of these
quantities to the atmospheric parameters.
In practise, the radial velocity and \vsini\ of the observed spectrum are obtained from the cross correlation function computed between
the observed spectrum and the synthesised one with parameters $\vec{\theta^c}$ and $\vsini = 0\,\, \kms$.
This cross correlation function is given by 
\begin{equation}
   \label{ccf1}
   CCF(v,0) = \int M_\lambda F_\lambda S_{\lambda'} (\vec{\theta^c},0)\, d\lambda,
\end{equation}
where $\lambda'$ is the Doppler shifted wavelength by a velocity $v$, given in the non-relativistic regime
 by $\lambda'=\lambda+\lambda v/c$, where $c$ is the speed of light.
A Gaussian function is fitted to the CCF and the mean of the Gaussian is assumed as the radial velocity of
the observed spectrum while the \vsini\ is determined from the full with at half maximum (FWHM) of the CCF peak as follows.
New CCFs are computed between the synthetic spectrum without rotation and the same synthetic spectrum
degraded by different amounts of \vsini:
\begin{equation}
   CCF(v,\vsini) = \int M_\lambda S_\lambda(\vec{\theta^c},\vsini) S_{\lambda'} (\vec{\theta^c},0) d\lambda
\end{equation}
The FWHM is computed for each CCF peak and a cubic spline is fitted to the relation between the
FWHM and \vsini\ values. This cubic spline is then used to find the \vsini\ of the observed spectra
from the FWHM of the CCF computed in equation~\ref{ccf1}.

In the next \zaspe\ iteration all the synthesised spectra are degraded to the \vsini\ obtained in the previous iteration, and
the observed spectrum is corrected in radial velocity by the amount found from the cross correlation function.
The degradation of the spectrum by rotation is performed with a rotational kernel computed following eq. 18.11
of \cite{gray:2008}. The limb darkening is taken into account using the quadratic limb-darkening law with coefficients
for the appropriate stellar parameters calculated using the code from \citet{espinoza:2015}.
The \vsini\ value for the first \zaspe\ iteration is obtained by cross-correlating the observed spectrum against one with
stellar parameters similar to those of the Sun.

\subsection{Grid exploration}
\label{exp}

The synthesis of high resolution spectra is a computationally intensive process. For this
reason \zaspe\ uses a pre-computed grid of synthetic spectra and, in order to obtain a synthetic
spectrum for an arbitrary set of stellar parameters, a cubic multidimensional interpolation is performed.

Given the known correlations between the three atmospheric parameters and the possibility of existence of
secondary minima in $X^2$ space due to the imperfect modelling of the synthetic spectra,
the approach of \zaspe\ for finding the global $X^2$ minimum is to
explore the complete parameter space covered by the grid and not to rely on slope minimisation techniques
that require an initial set of guess parameters. 

In each \zaspe\ iteration the extension and spacing of the of the parameter grid being explored changes.
In the first iteration \zaspe\ explores the complete atmospheric parameter grid with coarse spacing, while from the fourth iteration on, \zaspe\ starts focusing on smaller
regions of parameter space around $\vec{\theta^c}$ which are densely explored.
Table~\ref{grid} shows the extension and spacings that \zaspe\ uses for each iteration in its default
version, but these values can be easily modified by the user.

\zaspe\ terminates the iterative process when the parameters obtained after each iteration
do not change by significant amounts. In detail, convergence is assumed to be reached when
the parameters obtained in the $i$-th iteration do not differ by more than 10 K, 0.03 dex and 0.01 dex
in \teff, \logg\ and \feh, respectively, from the ones obtained in the $(i-1)$-th iteration.
This convergence is usually achieved after $\sim 5-10$ iterations.

\subsection{Parameter uncertainties and correlations}
\label{errors}
\begin{figure}
 \includegraphics[width=\columnwidth]{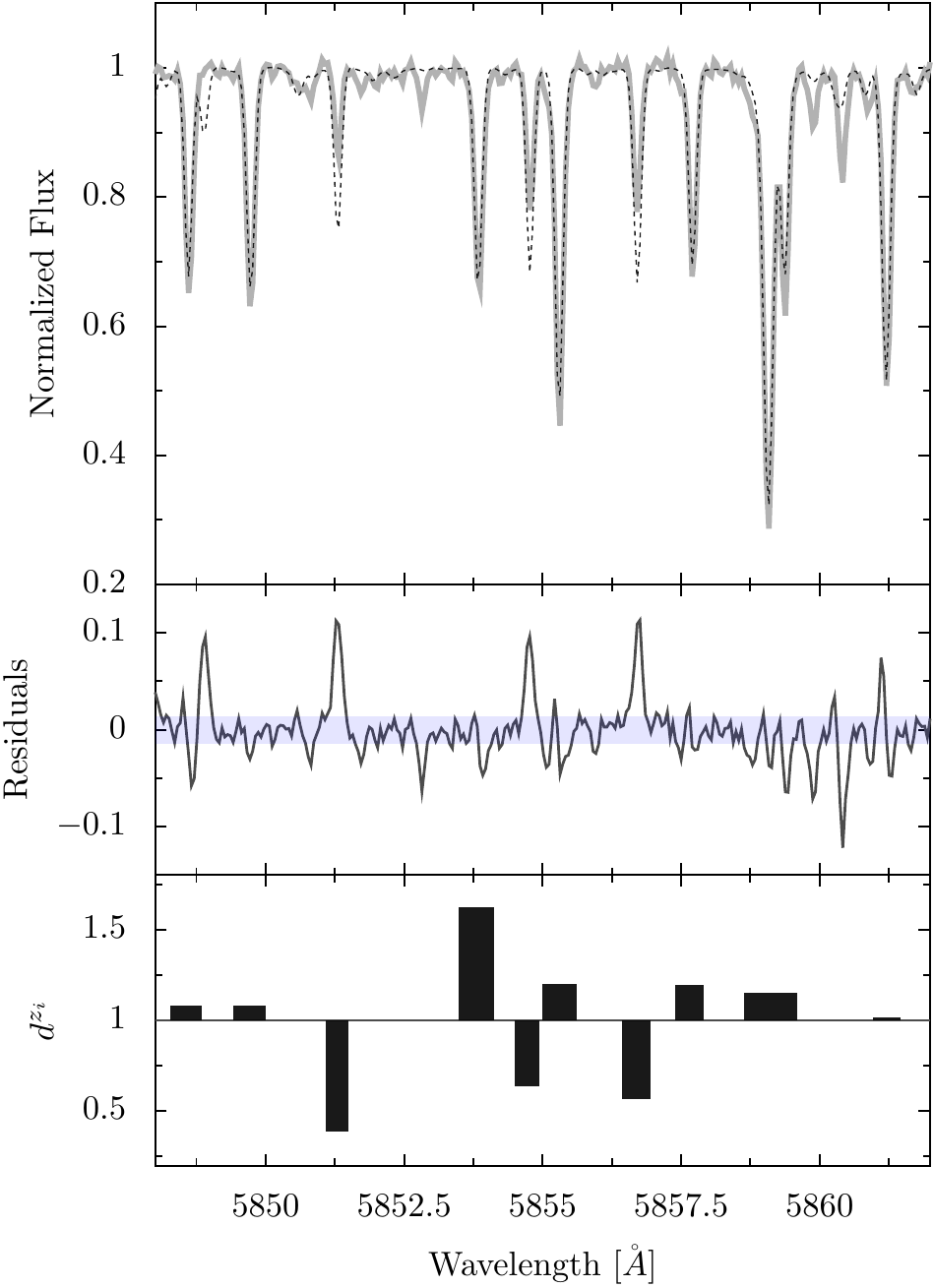}
 \caption{{\em Top}: portion of a high resolution echelle spectrum of a star (continuous line) and the synthetic
 spectrum that produces the best match with the data (dashed line). {\em Centre}: residuals between the two spectra and
 the expected 3$\sigma$ errors.
 Both panels show that the synthetic spectrum that best fits the data produces systematic mismatches in the zones
 of the absorption lines and that the errors are greater than the ones expected from the received flux. 
 {\em Bottom}: mismatch factors $d^{z_i}$ computed in the case of the 10 sensitive zones identified in this portion of the spectrum}
 \label{comp}
\end{figure}

As we mentioned in \S~\ref{sec:intro}, one major issue of the algorithms that use spectral synthesis
methods for estimating the stellar atmospheric parameters is the problem of obtaining
reliable estimates of the uncertainties in the parameters and their covariances.
\zaspe\ deals with this problem by assuming that the principal source of errors is the systematic
mismatch between the observed spectrum and the synthetic one.
The top panel of Figure~\ref{comp} shows a portion of
a high resolution spectrum of a star and the synthetic spectrum that produces the best
match with the data. Even though each absorption line is present in both spectra, the depth
of the lines is frequently different. This systematic mismatch can be further identified in the
central panel of Figure~\ref{comp}, where the residuals in the regions of the absorption lines
can be seen to be in several cases significantly greater than those expected just from photon noise.
In addition, the residuals are clearly non-Gaussian and highly correlated in wavelength.

\begin{figure}
 \includegraphics[width=\columnwidth]{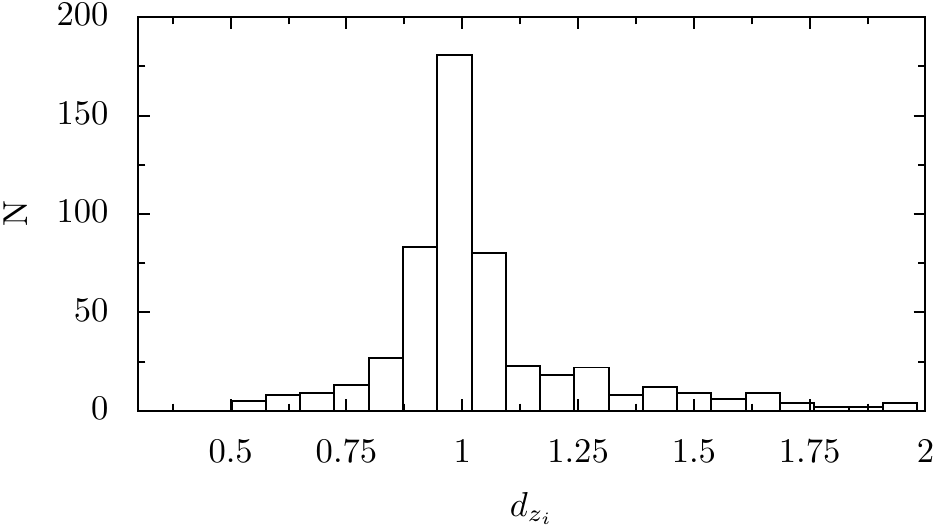}
 \caption{Histogram of the mismatch factors in the sensitive zones. In several regions of the spectrum
 the absorption lines of the synthetic spectrum can strongly deviate from the ones of the observed one.}
 \label{diet}
\end{figure}

Our approach to take into account the systematic mismatches, which builds upon the approach of \citet{grunhut:2009} is to define a random variable $D_i$ which is responsible for modifying the strength of each absorption feature in a sensitive zone $z_i$ of the synthesised
spectrum. If $S'^{z_i}_{\lambda}$ is a perfect synthetic spectrum in the $i$-th sensitive zone $z_i$, given a probability density $P(D)$ for the random variable $D$, an imperfect synthetic spectrum $S^{z_i}_{\lambda}$ (like the ones of the spectral libraries that \zaspe\ uses) is modeled as 
\begin{equation}
   S^{z_i}_{\lambda} = (S'^{z_i}_{\lambda} - 1) D + 1.
\end{equation}
An estimate of the probability density function $P(D)$ can be obtained
from the data itself by computing the set of mismatch factors, $d{z_{i}}$
for all sensitive zones, computed from the difference between the data and the
optimal synthetic spectrum found in the final \zaspe\ iteration. For each sensitive zone,
these factors $d^{z_i}$ are obtained from the median value, over all pixels in $z_i$, of the
division between the observed and synthetic spectra:
\begin{equation}
   d^{z_i} = \text{median}  \left( \frac{F^{z_i}_{\lambda}-1}{S^{z_i}_{\lambda}-1} \right) .
\end{equation}
The bottom panel of Figure~\ref{comp} shows the mismatch factors in the case of the
10 sensitive zones identified in that portion of the spectrum.
Figure~\ref{diet} shows an histogram of the mismatch factors for the same spectrum
of Figure~\ref{comp} but for a greater wavelength coverage ($5000 \AA < \lambda < 6000 \AA$).
The distribution of mismatch factors is pretty symmetric, centred around $d^{z_i}=1$ and shows
a wide spread of values. Most of the absorption lines of the synthetic spectrum that best fits the data
have values between 50\% and 200\% of the strength of the observed ones. Some lines can deviate even more,
however, these zones are rejected as strong outliers by \zaspe, as explained in \S~\ref{sec:zones}.

\zaspe\ estimates the probability distribution of the stellar atmospheric parameters by
running a random sampling method where a synthetic spectrum that produces the minimum $X^2$
is searched again a number $B$ of realizations, in the same way as described in the previous sections,
but using a modified set of model spectra in each realization. The only difference between the minimization
run on each realization and the original search is that the set of sensitive zones $\{z_i\}$ is kept fixed at the
set that \zaspe\ converged to. In each replication,  the strength of the lines of the synthetic spectra are
modified  by randomly selecting mismatch factors from the $\{d^{z_i}\}$ set, with replacement.
Each sensitive zone is modified by a different factor which can be repeated, but
the same factor is applied in each zone for the whole set of synthesised spectra.
In the random sampling method, the quantity that is minimised on each iteration $b$ is
\begin{equation}
\label{chism2}
   X^2_b = \sum_{\lambda} M_{\lambda} ( F_{\lambda} - (S_{\lambda}(\vec{\theta}) - 1)D_{\lambda} + 1)^2
\end{equation}
where $D_\lambda$ is a mask defined for each realisation and contains the mismatch factors for each sensitive zone.
In order to avoid possible biases in the final distribution of the parameters originating from the asymmetry in the
sampling function, when a factor is selected from $\{d{z_i}\}$ we include a 0.5 probability for this factor to 
take its reciprocal value, enforcing in practice symmetry in the function from which the factors are sampled.
After each realisation of the sampling method a new set of atmospheric parameters is found.
From these set of possible outcomes, the complete covariance matrix of the atmospheric
parameters can be estimated. After testing the method on spectra with different stellar atmospheric
parameters we found that about $B=100$ realisations are enough to obtain reliable parameter covariance matrices.

The procedure that \zaspe\ uses to obtain the errors and correlations assumes that the systematic mismatches
between the different zones are uncorrelated. This simplification of the problem means that some
systematic errors between the data and the models are not accounted for by our method. For example, if the abundance of
one particular atomic species strongly deviates from the one assumed in our model, the degree of mismatch
of the absorption lines of that element will be correlated. However, in \S~\ref{ssec:results} we will find that
our assumption is able to account for the typical value of systematic errors
in atmospheric parameters, as inferred from measuring the parameters with different methods.

\subsection{The reference spectral synthetic library}
\label{grids}

In order to determine the atmospheric stellar parameters of a star, \zaspe\ compares
the observed spectrum against a grid of synthetic models. In principle, after some
minor specifications about the particular format of the grid, \zaspe\ can use any pre-calculated grid.
We have tested \zaspe\ with two publicly available grids of synthetic spectra:
the one of \cite[hereafter C05]{coelho:2005}, which are based in the ATLAS model
atmospheres \citep{kurucz:1993}; and the one presented in \cite[hereafter H13]{husser:2013}, which is based
in the Phoenix model atmospheres. We have found that
both grids present important biases when comparing the
stellar parameters obtained using them with \zaspe\ for a set of reference stars.
In Figure~\ref{zaspe-comp} we show the comparison of the results obtained by \zaspe\ 
against the values presented in SWEET-Cat \citep{santos:2013} for a set of publicly
available spectra in the ESO archive.

SWEET-Cat is a catalogue of atmospheric stellar parameters of planetary host stars.
The parameters were computed using the equivalent width method and the ATLAS
plane-parallel model atmospheres \citep{kurucz:1993} on a set of high signal to noise and
high spectral resolution echelle spectra. We decided to use SWEET-Cat for benchmarking
our method because: (1) it includes stars with a wide range of stellar parameters; (2) the same
homogeneous analysis is applied to each spectrum; (3) the equivalent width method has
clear physical foundations and does not produce strong correlations between the inferred parameters;
and (4) the inferred parameters have been shown to be consistent with results obtained with different,
less model-dependent methods (infrared flux, interferometry, stellar density computed from transit
light-curve modelling) and also with standard spectral synthesis tools like \texttt{SPC} and \texttt{SME}
\citep{torres:2012}.

The top panels of Figure~\ref{zaspe-comp} show the comparison of the results
obtained by ZASPE using the H13 library.
These results deviate strongly from the reference values for the three
atmospheric parameters. The parameters are systematically underestimated
by 300 K, 0.6 dex and 0.3 dex on average in \teff, \logg\ and \feh, respectively.
There also appear to be quadratic trends in \teff\ and \logg\ which produce
greater deviations for hot and/or  giant stars. These systematic trends can be
expected from this kind of grid of synthetic spectra because the parameters of
the atomic transitions come from theory or from laboratory experiments, and are
not empirically calibrated with observed spectra.

Another possible source for these strong biases can be related to the different
model atmospheres used. We have estimated the atmospheric parameters of
the Sun with \zaspe+H13 finding that they present important deviations
with respect to the accepted reference values ($T_{eff \odot}^{H13}$=5430 K,
log$g_{\odot}^{H13}$=4.1 dex, [Fe/H]$_{\odot}^{H13}$=-0.3 dex).
These results show that if the strong observed biases are produced due to the use of
different model atmospheres, the PHOENIX models are less precise than the ATLAS ones
for estimating atmospheric parameters. 

The central panels of Figure~\ref{zaspe-comp} correspond to the results obtained
by ZASPE using the C05 library.
Even though the average values determined with the C05 grid are more compatible with the
reference values than the ones obtained with the H13 grid, there is
a strong trend in $\Delta\teff$. The systematic trend tends to bring the
values of \teff\ towards the one of the Sun ($\approx$5750K) and can produce
deviations of $\approx$500 K for F-type stars.
In this case both set of results are obtained using the same model atmospheres.
The origin of the observed bias is unknown, but it can be plausibly related to two procedures that were
adopted in the generation of the C05 grid. First, the oscillator strengths (log$gf$) of several
\ion{Fe}{}\ transitions were calibrated using a high resolution spectrum of the Sun, which could bias
the results if the physical processes responsible for the formation of the lines are not accurately
modelled by the synthesising program; and second, all the spectra with $\logg>3.0$ were
synthesised assuming a solar micro turbulence value of $v_t=1.0$ \kms, but FGK-dwarfs
 have measured microturbulence values in the range of $\approx$0-6 \kms.
The behaviour obtained for the values of the other parameters show less biases. However,
the trend in \teff\ coupled with the correlations in the atmospheric parameters
induce an important dispersion in \logg\ and \feh.

\subsection{A new synthetic grid}
\label{ngrid}

As shown in the last section, it is not straightforward to use public libraries of synthetic spectra
for estimating atmospheric parameters of stars due to strong systematic trends and biases
that can arise due to erroneous physical assumptions and calibrations.
For that reason we decided to synthesise a new grid. We used the \texttt{spectrum}
code \citep{gray:1999} and the Kurucz model atmospheres \citep{castelli:2004} with solar scaled abundances.
In order to avoid biases in \teff\ related to assuming a fixed microturbulence value we 
assume that the microturbulence is a function of \teff\ and \logg. \cite{ramirez:2013}
established an empirical calibration of the microturbulence  as a function of the three atmospheric
parameters but the validity of the proposed relation was limited to stars having $\teff>5000$ K. We thus decided to
base our microturbulence calibration on the values computed in SWEET-Cat by \cite{santos:2013}. We considered only
the systems having the homogeneity flag and by visually inspecting the dependence of the microturbulence with respect to
the atmospheric parameters, we defined three different regimes for our empirical micro turbulence law.
For dwarf stars ($\logg>3.5$) the microturbulence was assumed to depend on \teff\ by a third degree polynomial,
while for sub-dwarf and giant stars the microturbulence was fixed to two different values as follows

\begin{figure*}
 \includegraphics[width=16cm]{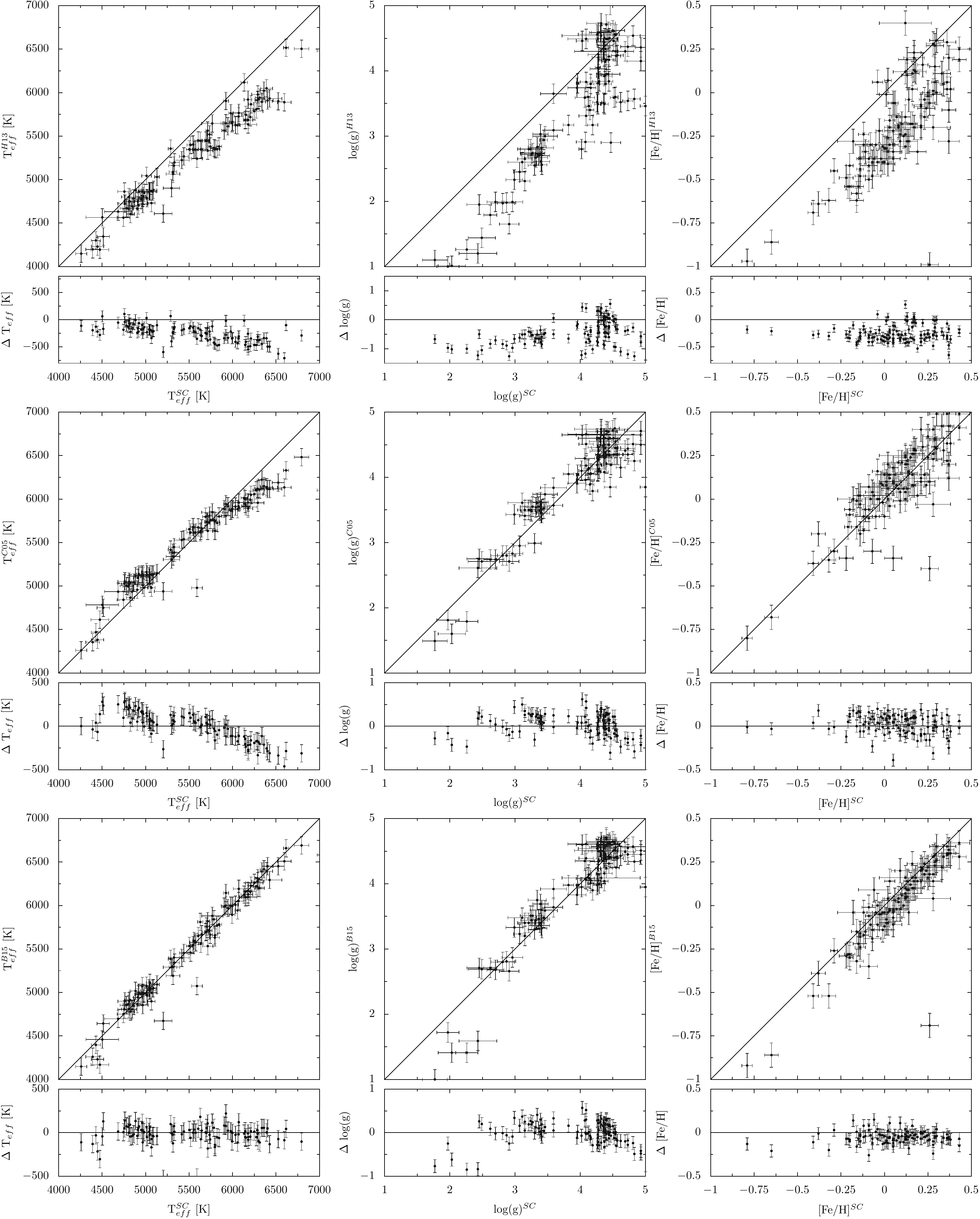}
 \caption{Comparison of the atmospheric parameters obtained by \zaspe\ using three different libraries
 of synthetic spectra against the values reported in SWEET-Cat. The top panels correspond to the results obtained
 using the H13 grid, where strong biases and systematic trends are present in the three parameters
 probably because the parameters of the atomic transitions were not empirically calibrated. The central panels correspond
 to the results obtained using the C05 grid, where a strong systematic trend in \teff\ drives \teff\ values towards
 that of the Sun. The bottom panels show the results obtained by \zaspe\ when using the synthetic library presented in
 this work. Results are compatible with the values reported in SWEET-Cat and no strong systematic trends can be identified.}
 \label{zaspe-comp}
\end{figure*}

\begin{align*}
 v_t & = -36.125 + 0.019 \teff\  &&\\
 & -3.65 \times 10^{-6} \teff^2 + 2.28 \times 10^{-10} \teff^3 &&
      (\logg>3.5)\\
 v_t & =  1.2 \kms\  && (3.0<\logg<3.5)\\
 v_t &=1.6 \kms && (\logg<3.0)\\
\end{align*}

We used the line list provided in the \texttt{spectrum} code. We initially synthesised a grid of spectra using the original parameters of the transitions
in the line list. However, after testing the grid with \zaspe\ we found that while the estimated \teff\ and \feh\ values were closer to the Sweet-Cat ones
than the values found using the other two public libraries, some slight but significant biases in these parameters were still present and also the \logg\ values were strongly underestimated by $\sim 0.8$ dex. For this reason we decided to perform a similar approach to C05, and  
we tuned the log$gf$ of several ($\sim$400) prominent atomic lines. As opposed to C05, though, we did not use the spectrum of the sun to perform
the tuning, but instead we used the spectra of a set of stars that have some of their atmospherical parameters obtained using more direct procedures.
In particular we used stars whose \teff\ were measured by long baseline interferometry \citep{boyajian:2012,boyajian:2013} and another set of 
stars with \logg\ values precisely determined through asteroseismology using $Kepler$ data \citep{silva:2015}. For the latter sample of stars we obtained their spectra from the public Keck/HIRES archive, while for the former sample we obtained the spectra from the same archive but we also use data of the FEROS spectrograph that was found in the ESO archive. Tables~2 and 3 show the stars that were used to adjust the log$gf$ values.
For each absorption line we determined the best log$gf$ value in the case of each reference spectrum by building synthetic spectra in this particular spectral region with different values of log$gf$ but with the stellar parameters fixed to the ones obtained by asteroseismology or interferometry.
For each star we found the synthetic spectrum that produces the smaller $\chi^2$ and we save the log$gf$ value of that model. Then we used the
median value of the log$gf$ determined for the different stars as the calibrated log$gf$ value of the particular atomic transition.

In addition to the  log$gf$ values of the $\sim400$ spectral lines,we also manually adjusted the damping constants of the \ion{Mg}{I}b triplet and \ion{Na}{I} doublet using a similar procedure.
\texttt{spectrum} uses the classical van der Wals formulation to generate the wings of the strong lines but
this procedure has been found to underestimate the strength of the absorption features. A common solution is to
include an enhancement factor to correct for this behaviour. In our case, we determined this empirical adjustment factor for
each of these strong lines using the above mentioned set of standard stars.  We found that the adjustment factor has a
temperature dependence. \cite{anstee:1991} developed a detailed approximation of the van der Waals theory in which the
temperature dependence of the damping constant was determined to follow a power law. In our case, we empirically treated the temperature
dependence of the damping constants by fitting linear relations to the enhancement factors determined from the standard
stars as a  function of the temperature for each strong line. These parameters were then used to synthesise the \ion{Mg}{I}b and \ion{Na}{I}
lines for spectra with different values of \teff.

The spectral range of our grid goes from 4900\AA\ to 6100\AA. This range was selected because
most of the spectral transitions for FGK-type stars are located at shorter wavelengths than  6000 \AA\ but for
$\lambda < 5000 \AA$ spectral lines become excessively crowded which complicates the process of adjusting
the log$gf$ values. The grid limits and spacings of the stellar parameters of the grid we synthesised are
\begin{itemize}
      \item \teff: 4000K --- 7000K, $\Delta\teff$=200K
      \item \logg: 1.0 dex --- 5.0 dex, $\Delta\logg$= 0.5 dex
      \item \feh: -1.0 dex --- 0.5 dex, $\Delta\feh$=0.25 dex.
\end{itemize}
We used a multidimensional cubic spline to generate the model atmospheres with
atmospheric parameters not available in the original set of atmospheres provided by the Kurucz models.

\begin{figure*}
 \includegraphics[width=\textwidth]{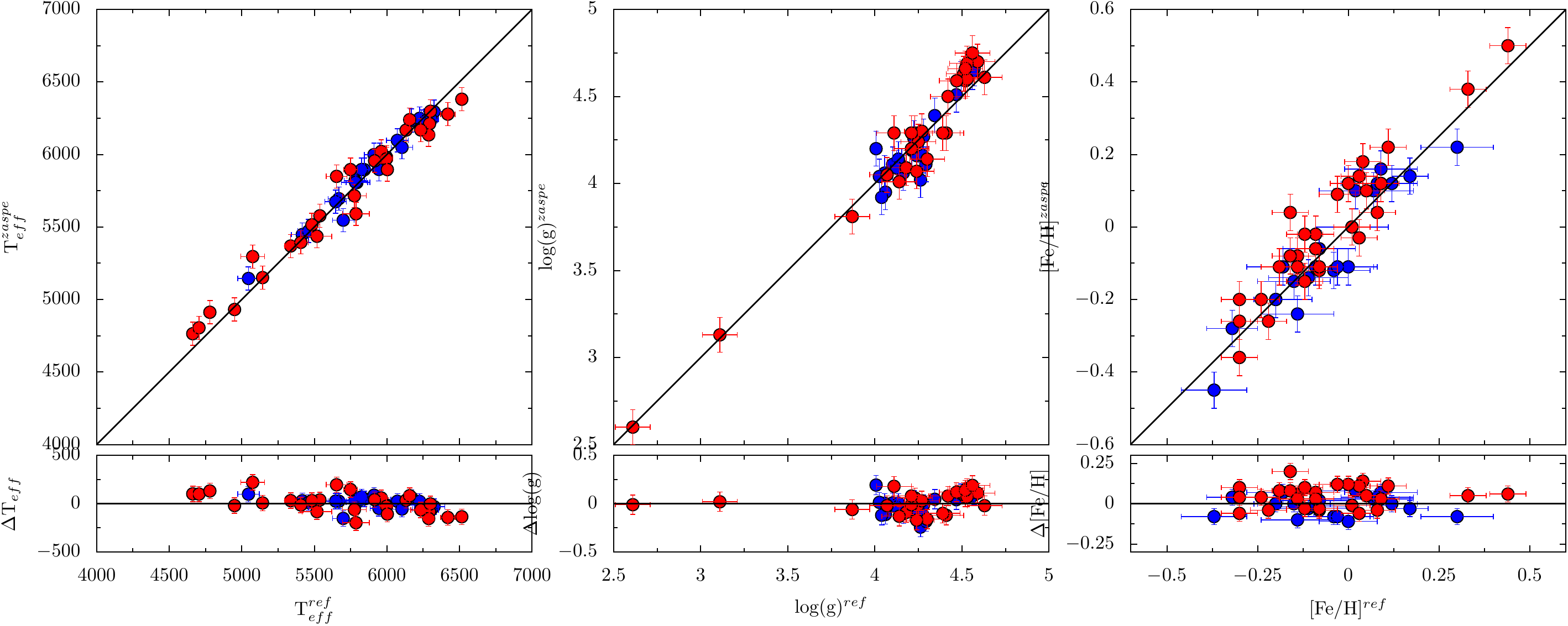}
 \caption{ Comparison between the parameters obtained by \zaspe\ using the new grid and the reference values for the sets of stars with asteroseismological (blue) and interferometric (red) derived parameters. The left panel shows the results in the case of \teff\ and the right panel for \logg.}
 \label{inter-aste}
\end{figure*}

\begin{figure*}
 \includegraphics[width=\textwidth]{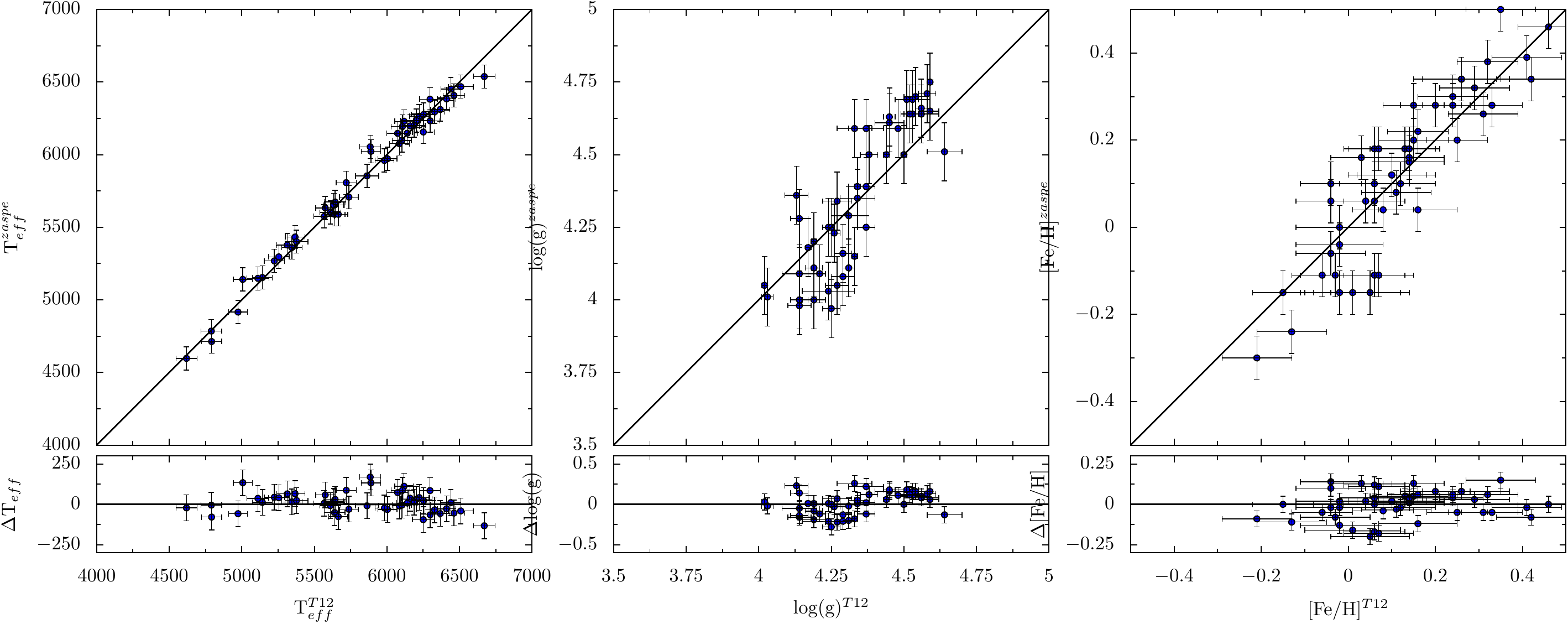}
 \caption{Comparison between the parameters obtained by \zaspe\  using the new grid and the reference values for the set of stars analysed in \citet{torres:2012}. Left, central and right panels correspond to the comparisons in \teff , \logg\  and \feh , respectively}
 \label{torres}
\end{figure*}

The bottom panels of Figure \ref{zaspe-comp} show the results obtained using \zaspe\ with
this new grid of synthetic spectra against the values stated in SWEET-Cat. The results agree very
well with the reference values and no evident trends are present.
The \teff\ shows an excellent agreement with only two outliers at present. The results obtained for \logg\
have some tentative systematic trends. In particular, we note that SWEET-Cat report some \logg\ values greater
than 4.7 dex, but we note that surface gravities higher than that are not common for FGK-type stars so those values are suspect.
The \feh\ values present no offset trends, but a systematic bias can be identified. \feh\ values are on average
underestimated by 0.05 dex as compared to the SWEET-Cat values. However, differences of $\approx$0.09 dex in
\feh\ have been previously reported when comparing SWEET-Cat metallicities against the
ones computed with the ones obtained via spectral synthesis techniques, so the offset we observe is within the expected range given the different techniques used \citep{mortier:2013}.
In order to further check the performance of our new grid we used other three different samples of stars with stellar parameters
obtained in an homogeneous way. First we used our two sets of stars with stellar parameters obtained using
interferometry and asteroseismology which are shown in Tables~2 and 3. Figure~\ref{inter-aste} shows the comparison of the parameters obtained
using \zaspe\ with our new grid as function of the reference values.
The third sample of stars that we use corresponds to the exoplanet host stars analysed by \cite{torres:2012}, where the \logg\ values
where precisely obtained by using the measured stellar densities obtained from the transit light curve. Figure~\ref{torres} shows the
results obtained by \zaspe\ as function of the parameters found in the mentioned study.
For these three samples of stars, the parameters obtained with \zaspe\ using our new grid are in good agreement with the reference values.
However, there is still a slight but significant overestimation of the \logg\ values in the case of dwarf stars. This problem can be produced
because (i) not all the spectral lines were empirically calibrated, and (ii) we are imposing that the modelling errors are originated from
unreliable log$gf$ values, and therefore, when we perform the calibration that significantly improves the quality of the synthetic grid,
some additional weaker systematic errors could be introduced.

Our new spectral library has been made publicly available\footnote{http://www.astro.puc.cl/$\sim$rbrahm/new\_grid.tar.gz}.

\begin{table}
\label{inter}
 \centering
  \caption{Sample of stars with temperatures measured using interferometric observations \citep{boyajian:2012,boyajian:2013}  that were used
  to empirically calibrate log$gf$ values and damping constants of prominent absorption lines.}
  \begin{tabular}{@{}lccccccc@{}}
  \hline
   
   Name     &   \teff\ [K] &  $\sigma_{\teff}$  [K] &  \logg  &  \feh & Instrument\\
 \hline
GJ105       & 4662 & 17 & 4.52 & -0.08 & FEROS\\
GJ166A     & 5143 & 14 & 4.54 & -0.24 & FEROS \\
GJ631       & 5337 & 41 & 4.59 & 0.04 & FEROS \\
GJ702A     & 5407 & 52 & 4.53 & 0.03 & FEROS \\
HD102870 & 6132 & 26 & 4.11 &  0.11 & FEROS \\
HD107383 & 4705 & 24 & 2.61 & -0.30  & HIRES\\
HD109358 & 5653 & 72 & 4.27 & -0.30  & HIRES\\
HD115617 & 5538 & 13 & 4.42 &  0.01 & FEROS \\
HD131156 & 5483 & 32 & 4.51 & -0.14  & FEROS\\
HD142860 & 6294 & 29 & 4.18 & -0.19 & FEROS \\
HD145675 & 5518 & 102 & 4.52 & 0.44  & HIRES\\
HD1461     & 5386 & 60 & 4.20 &  0.16 & FEROS \\
HD146233 & 5433 & 69 & 4.25 & -0.02 & FEROS \\
HD16895   & 6157 & 37 & 4.25 & -0.12  & HIRES\\
HD182572 & 5787 & 92 & 4.23 &  0.33  & FEROS\\
HD19373   & 5915 & 29 &4.21 & 0.09  & HIRES\\
HD20630   & 5776 & 81 & 4.53 &  0.0 & FEROS \\
HD210702 & 4780 & 18  & 3.11 & 0.03  & HIRES\\
HD222368 & 6288 & 37 & 3.98 & -0.08 & FEROS \\
HD22484   & 5997 & 44 & 4.07 & -0.09  & FEROS\\
HD30652   & 6516 & 19 & 4.30 & -0.03 & FEROS \\
HD33564  &  6420 & 50 & 4.24 & 0.08 & HIRES \\
HD34411  &  5749 & 48 & 4.21 & 0.05  & HIRES\\
HD39587   & 5961 & 36 & 4.47 & -0.16 & FEROS\\
HD4614    & 6003 & 24 & 4.39 & -0.30  & HIRES\\
HD4628     & 4950 & 14 & 4.63 & -0.22 & FEROS \\
HD7924     & 5075 & 83 &  4.56 & -0.14  & HIRES\\
HD82328   & 6300 & 33 &  3.87 & -0.12  & HIRES\\
HD82885   & 5434 & 45 &  4.39 & 0.06 & HIRES \\
HD86728	  & 5612 & 52 &  4.26 & 0.20 & HIRES \\
HD90839   & 6233 & 68 & 4.41 & -0.16 & HIRES \\
\hline
\end{tabular}
\end{table}

\begin{table*}
\label{aste}
 \centering
  \caption{Sample of stars with \logg\ values measured using asteroseismology of $Kepler$ data \citep{silva:2015}, that were used
  to empirically calibrate log$gf$ values and damping constants of prominent absorption lines.}
  \begin{tabular}{@{}lccccccccc@{}}
  \hline
   
   Name     &   \teff\ [K] &  $\sigma_{\teff}$  [K] &  \logg & $\sigma_{\logg}$  &   \feh & $\sigma_{\feh}$ & Instrument\\
 \hline
KOI 1612 &   	6104 &  74 &  4.293 &  0.004 &  -0.20 &  0.10 & HIRES \\
KOI 108 &   	5845 &  88 &  4.155 &  0.004 &   0.07 &  0.11 & HIRES \\
KOI 122 &   	5699 &  74 &  4.163 &  0.003 &   0.30 &  0.10 & HIRES \\
KOI 41 &   	5825 &  75 &  4.125 &  0.004 &   0.02 &  0.10 & HIRES \\
KOI 274 &   	6072 &  75 &  4.056 &  0.013 &  -0.09 &  0.10 & HIRES \\
HIP94931 &  5046 &  74 &  4.560 &  0.003 &  -0.37 &  0.09 & HIRES \\
KOI 246 &    	5793 &  74 &  4.280 &  0.003 &   0.12 &  0.07 & HIRES \\
KOI 244 &    6270  & 79  & 4.275  & 0.008  & -0.04  & 0.10  & HIRES \\
KOI 72 &    	5647 &  74 &  4.344 &  0.003 &  -0.15 &  0.10 & HIRES \\
KOI 262 &    	6225 &  75 &  4.135 &  0.008 &  -0.00 &  0.08 & HIRES \\
KOI 277&     5911 & 66 & 4.039 & 0.004 & -0.20 & 0.06 & HIRES \\
KOI 123 &    5952  & 75  & 4.213  & 0.008  & -0.08  & 0.10  & HIRES \\
KOI 260 &    6239  & 94  & 4.240  & 0.008  & -0.14  & 0.10  & HIRES \\
KOI 1925 &    	5460 &  75 &  4.495 &  0.002 &   0.08 &  0.10 & HIRES \\
KOI 5&    5945 & 60 & 4.007 & 0.003 &  0.17 & 0.05  & HIRES \\
KOI 245 &    	5417 &  75 &  4.570 &  0.003 &  -0.32 &  0.07 & HIRES \\
KOI 7 &    	5781 &  76 &  4.102 &  0.005 &   0.09 &  0.10 & HIRES \\
KOI 263 &    5784  & 98  & 4.061  & 0.004  & -0.11  & 0.11   & HIRES \\
KOI 975 &    	6305 &  50 &  4.026 &  0.004 &  -0.03 &  0.10 & HIRES \\
KOI 69  &   5669   & 75   &4.468   &0.003   &-0.18   & 0.10   & HIRES \\
KOI 42 &   6325  & 75  &4.262  &0.008  & 0.01  & 0.10   & HIRES \\
\hline
\end{tabular}
\end{table*}

\section{Performance}
\label{ssec:results}
\begin{figure}
 \includegraphics[width=\columnwidth]{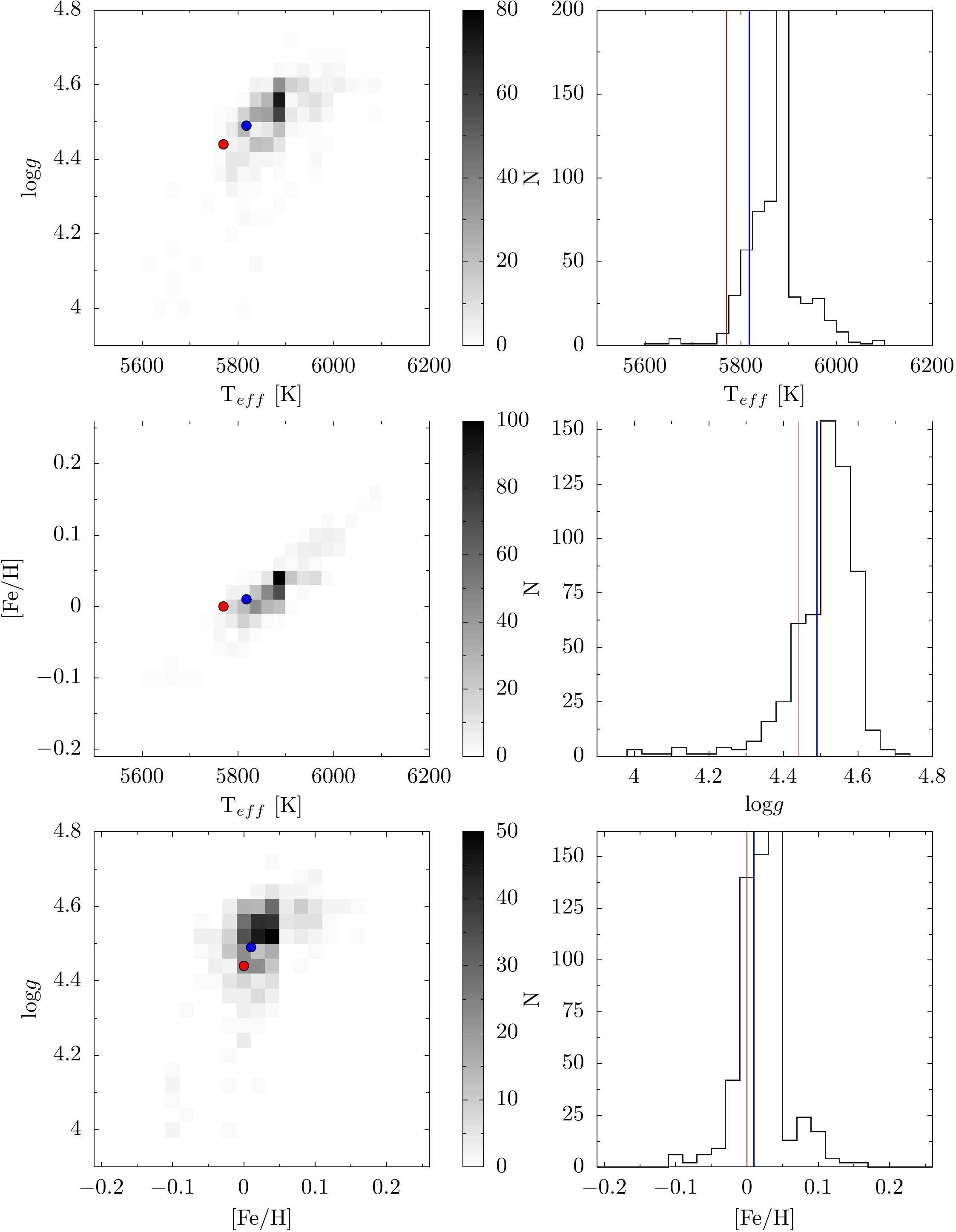}
 \caption{Results obtained by \zaspe\ for the Sun. The left panels show the distribution
 of possible sets of atmospheric parameters obtained from the random sampling method. Strong
 correlations between the parameters are found. The blue circles correspond to the parameters of the
 synthetic spectrum that produce the best match, while the red circles are reference values from the literature.
 The right panels correspond to the marginalised distributions of outcomes for each atmospheric parameter.
 Blue lines show the parameters of the synthetic spectrum that produce the best match, while red lines are the reference values.
 The errors reported by \zaspe\ correspond to the standard deviations of these distributions.
 }
 \label{sun}
\end{figure}

\begin{figure}
 \includegraphics[width=\columnwidth]{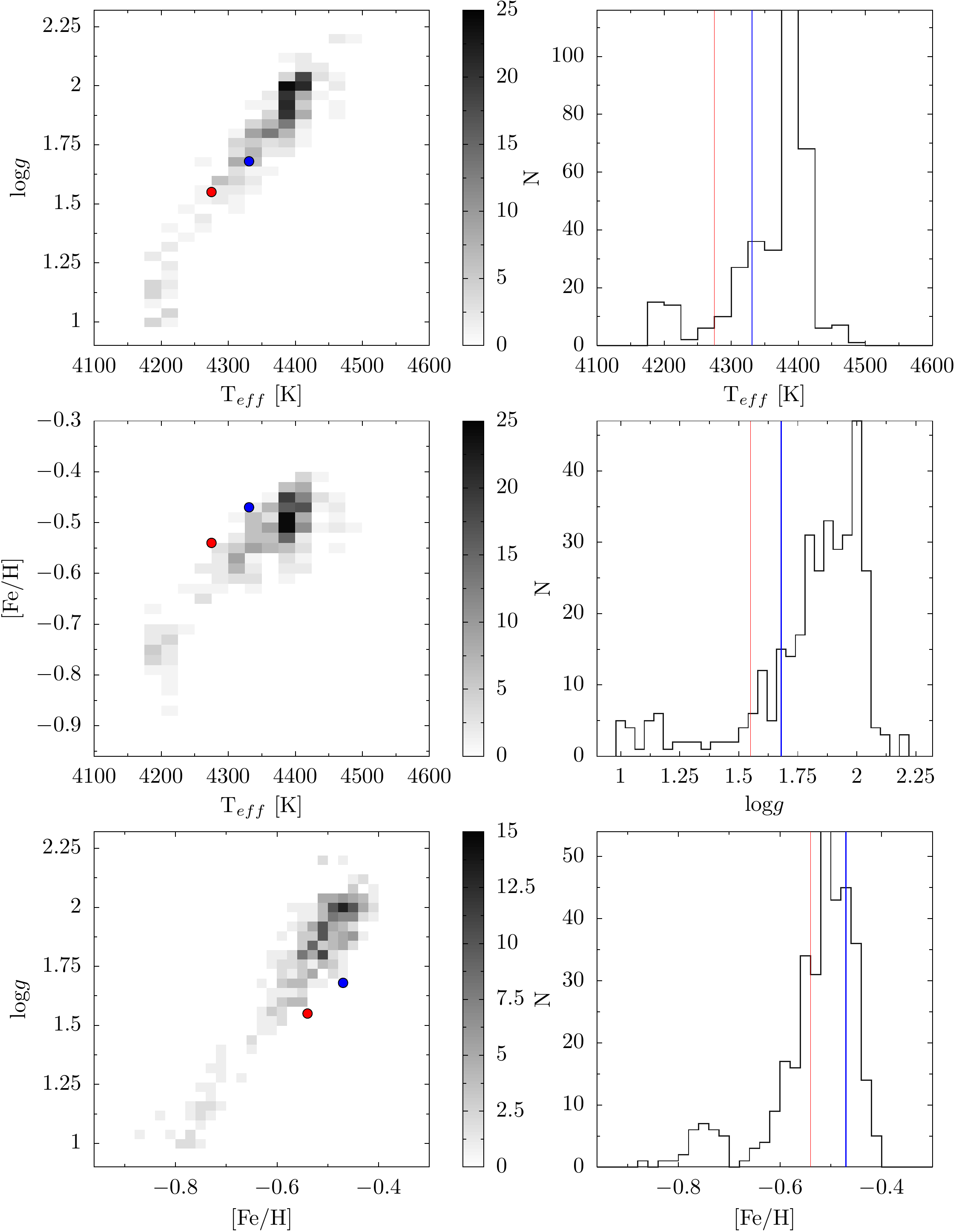}
 \caption{Same as Figure~\ref{sun} but for Arcturus.
 }
 \label{arcturus}
\end{figure}

 As an example of the performance of \zaspe, we present here the results we obtain when using it to
analyse the spectra of the Sun and Arcturus.
The spectra of these two objects have been studied extensively and they are used
commonly to calibrate and validate spectral studies of stars.
We obtained raw data from the ESO archive for both stars observed with the FEROS spectrograph \citep{kaufer:1998}.
We processed them through an automated reduction and extraction pipeline we have developed for FEROS and other echelle spectrographs \citep[Brahm et al. in prep]{jordan:2014}.
The grid of synthetic spectra used by \zaspe\ in this analysis was generated by us and is described in \S~\ref{ngrid}.
The spectral range selected for analysing the data was from 5000 to 6000 $\AA$,
which ensures a great amount of spectral transitions including the \MgI\ triplet, which is the most
pressure sensitive feature for dwarf stars. Figures~\ref{sun} and \ref{arcturus} summarize the results we obtain.
The left panels show Hess diagrams in various planes using the outcome of the random sampling realisations,
while the right panels show the marginalised distribution functions of the stellar atmospheric parameters.
The best fit parameters are marked with red circles in the panels on the left and vertical red lines in the ones on the right. Reference values of the stellar parameters are marked with blue circles in the panels on the left and blue lines in the ones on the right.

In the case of the sun the best fit parameters and errors we obtained were: \teff=5818 $\pm$ 59 K, \logg=4.49 $\pm$ 0.09 dex and
\feh=0.01 $\pm$ 0.04 dex. These results are compatible with the accepted parameters of the sun being 0.8$\sigma$, 0.6$\sigma$ and
0.3$\sigma$ apart in \teff, \logg\ and \feh, respectively. The results obtained for Arcturus were: \teff=4331 $\pm$  63 K, \logg=1.68 $\pm$ 0.25 dex and
\feh=-0.48 $\pm$ 0.09 dex. In Figure~\ref{arcturus} we include the parameters
computed by \cite{melendez:2003} which are compatible  with the results obtained by \zaspe\ at the 1$\sigma$ level.

In both cases, \zaspe\ shows there is a wide spread of possible outcomes which confirms the idea that the principal source of uncertainty is the imperfect modelling of the synthesised spectra. The uncertainties in the parameters that
\zaspe\ reports are computed from the standard deviation of the values obtained in the random sampling
simulations. The uncertainties in the parameters we found for the Sun are smaller than the ones found for Arcturus.
This serves to illustrate that the amplitude of the uncertainty in the atmospheric parameters varies with spectral type, and that the synthetic grid we used has a better calibration for dwarf stars than for giant stars. It is therefore not accurate to adopt universal minimum uncertainties, as is often done in the literature.

The left panels of Figures~\ref{sun} and \ref{arcturus} also show the existence of strong correlations between the parameters.
The Pearson correlation coefficients $\rho$ between the parameters in the case of the Sun are: $\rho_{\teff-\logg}=0.63$, $\rho_{\teff-\feh}=0.89$
and $\rho_{\logg-\feh}=0.49$. For Arcturus the correlations we found are $\rho_{\teff-\logg}=0.95$, $\rho_{\teff-\feh}=0.87$
and $\rho_{\logg-\feh}=0.92$.

One first thing to note about the performance of \zaspe\ on the Sun and Arcturus, beyond the fact that the resulting stellar parameters are consistent with known values produced by current state of the art analyses, is the magnitude of the uncertainties. Despite the very high signal-to-noise ratio of the spectra, the estimated uncertainties in \teff\ are $\sim 50$ K. This compares very well with the uncertainty of $\sigma_{\teff}=59$ K that \citet{torres:2012}  adds in quadrature to their formal uncertainties. This uncertainty is obtained 
from the overall scatter of their measurements  for stars with multiple determinations obtained with different methods (\texttt{SPC}, \texttt{SME} and/or \texttt{MOOG}). In the same vein, the uncertainties in \feh\ for the Sun\footnote{The \feh\ uncertainty for Arcturus is higher by a factor of $\approx 2$. This is a consequence of the less constrained value of \logg\ for a giant, which has an impact on the uncertainty of \feh.} are of order $\sim 0.05$ dex, compared with the value of $\sigma_{\feh}=0.062$ adopted by \citet{torres:2012}. From this exercise we can see that the uncertainties returned by \zaspe\ are a realistic reflection of the model uncertainties that dominate in our case. As opposed to the methods based on repeated measurements on a sample of objects, \zaspe\ can provide that uncertainty on a per spectrum basis, and also provides the correlation with other parameters. 

\begin{figure*}
 \includegraphics[width=\textwidth]{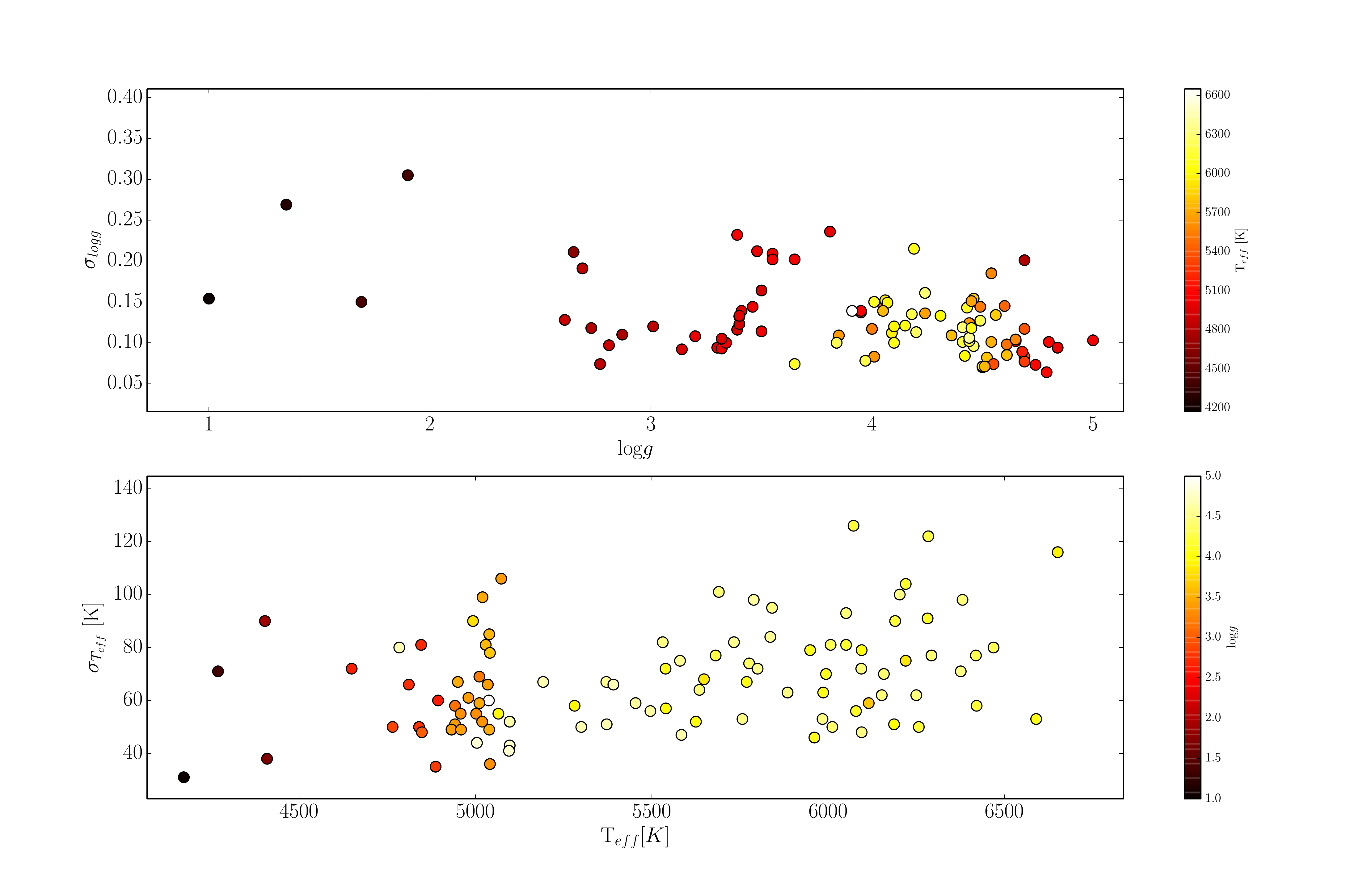}
 \caption{Top: errors in \logg\ reported by \zaspe\ as function of the \logg\ values. Dwarfs stars have in general smaller errors than giant stars.
 Bottom: errors in \teff\ reported by \zaspe\ as function of the \teff\ values. In the case of dwarf stars, hotter stars tend to have larger errors.
 }
 \label{sims}
\end{figure*}

In order to explore how the magnitude of the computed errors depend on the atmospheric parameters, we analysed the results that were obtained by \zaspe\ on the dataset presented in \S~\ref{ngrid}, where
we  obtained the atmospheric parameters and their associated errors for a set of FEROS spectra of stars that have been already analysed by SWEET-Cat.
From this sample we conclude that there is no strong dependence between the magnitude of the errors that we estimate and the atmospheric parameters of the star. However, there are two tentative trends
that are shown in Figure~\ref{sims}. The top panel of the figure shows that dwarf stars tend to have lower errors
in \logg\ than giant stars. The origin of this correlation can be associated with the tight pressure sensitivity of the shape of the wings of strong absorption lines, which is only present in dwarf stars. In the case of giant stars
the principal factor that produces variation in \logg\ are subtle changes in the depth of shallow lines generated from variations in the continuum absorption. The bottom panel of Figure~\ref{sims} shows that for dwarf stars,
the \teff\ errors computed by \zaspe\ tend to be higher or at least have a larger dispersion for hotter stars, which can arise from the higher rotational velocity F-type stars have in comparison to G-type stars, but
also because at higher temperatures (\teff\ $>$6000 K), a large fraction of the elements in the atmosphere start to get ionised and therefore there are less available absorption lines. However, the reported trends
of both panels show important levels of scatter. In particular there is a cluster of stars with $\teff\approx5000$ K, $\logg\approx$ 3.5 dex and similar \feh\ values that shows a large scatter in the magnitude of their errors.
The source of this dispersion may be associated to other systematic effects, like differences in particular abundances or incorrect assumptions in the micro- and macro-turbulence values.

As a further example of the performance of \zaspe, we analysed an archival Keck/HIRES \citep{vogt:1994} spectrum of WASP-14. We chose this star as it is representative of the use of atmospheric parameter estimation in the process of discovery and characterization of exoplanets, which was the main motivation for developing \zaspe. Additionally, WASP-14 was analysed with \texttt{Starfish} by \citet{czekala:2014}. As \texttt{Starfish} is the only other approach we are aware of that attempts to properly take into account the model uncertainties as we do in this work, it offers a very interesting point of comparison. \citet{czekala:2014} estimate the stellar parameters of WASP-14 using a spectrum from the TRES spectrograph on the Fred Lawrence Whipple Observatory 1.5 m telescope and fixing \logg\ to the value obtained by \citet{torres:2012} by fitting the transit light-curve, namely $\logg=4.29$. They estimate parameters using both Kurucz and PHOENIX stellar atmospheric models.
The \texttt{Starfish} estimates and their uncertainties are presented in their Table~1 and are (values using Kurucz models) $\teff=6426\pm21$ K, $\feh=-0.26\pm0.01$ and $\vsini=4.47\pm 0.06$ \kms. 

\begin{figure*}
 \includegraphics[width=\textwidth]{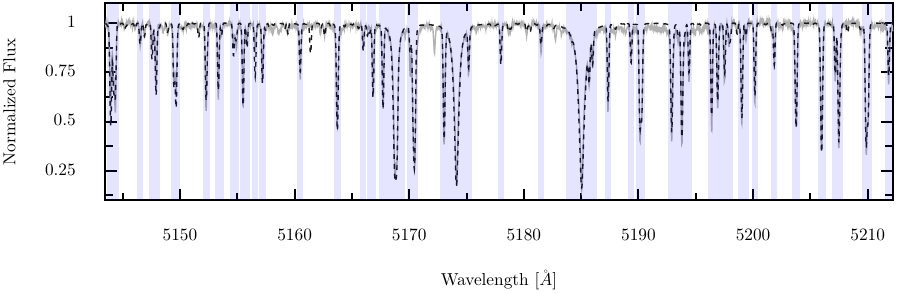}
 \caption{A portion of the spectrum of WASP-14 observed with Keck/HIRES in the zone of the \ion{Mg}{I}b triplet. The grey line corresponds to the observed spectrum, while the dashed line is the synthetic spectrum with atmospheric parameters determined by \zaspe. Shaded in blue are the sensitive zones that \zaspe\ selected.
 }
 \label{wasp14}
\end{figure*}

Running \zaspe\ on the Keck/HIRES spectrum with fixed $\logg=4.29$ results in the following estimates: $\teff=6515\pm64$ K, $\feh=-0.15\pm0.04$ and $\vsini=5.55\pm 0.37$ \kms. Again, the uncertainties are reasonable based on what is expected from studies that have obtained measurements with different methods such as \citet{torres:2012}, and are actually made somewhat artificially low by fully fixing \logg\footnote{The stellar parameters obtained leaving \logg\ free are: $\teff=6501\pm134$ K, $\feh=-0.17\pm0.07$, $\vsini=5.58\pm 0.5$ \kms and $\logg=4.22\pm0.18$.}. The difference with the uncertainties obtained by \texttt{Starfish} are substantial, with the \texttt{Starfish} uncertainties being underestimated based on the experience provided by studies such as that of \citet{torres:2012}. This is also clear from comparing the parameters derived by \texttt{Starfish} on the same data using different stellar models, as they are in some cases formally inconsistent given their error bars, something that should not be the case if the uncertainties arising from model imperfections have been properly estimated. Figure~\ref{wasp14} shows a portion of the HIRES spectrum of WASP-14 and the synthetic spectrum with the optimal parameters derived by \zaspe. The sensitive zones determined by \zaspe\ are shaded blue.

It is worthwhile trying to understand why the approach of \citet{czekala:2014} leads to underestimated uncertainties. Their approach is very principled, and being immersed in a likelihood, it is very appealing for inference. Their approach takes into account the mismatches between models and data through modelling the variance structure with a Gaussian process. To that effect, a mixture of non-stationary kernels that indicate regions of very strong deviation, and a stationary global kernel, are used. The non-stationary kernels, with large variances, have the effect of ignoring regions where those kernels are instantiated, and is a way of eliminating lines that are outliers in a principled way. The stationary kernel accounts for the typical mismatch between the model and the data, and it is chosen to be of the form of a Mat\'ern $\nu=3/2$ kernel, tapered by a Hann window function to keep the global covariance matrix sparse. In this approach, the possible mismatches between the model and the data are given by the space of functions generated by the Mat\'ern $\nu=3/2$ Kernel with the hyperparameter distributions learned in the inference process. 

The key observation is that the mismatches are not appropriately described by a stationary kernel, as they ought to exist mostly around the lines, and thus the process that would be needed to account for the mismatch structure is fundamentally non-stationary. In our re-sampling scheme, we just modify the depth of the lines, exploring thus systematically variations in the models that have physical plausibility. Variations given by a stationary Gaussian kernel will have no correlation with the line structure, and would be therefore mostly unphysical. This can be seen in the right panels of Figure~4 in \citet{czekala:2014}, the random draws from the stationary kernel have structure on locations that are uncorrelated with the spectral lines. The stationary kernel encapsulates the typical covariance structure of the mismatch, including large swaths of the spectrum that are continuum where little mismatch is observed, as those regions are less sensitive to the parameters. One expects then that the amplitude of the variance is a sort of average description of continuum and line regions, and would thus underestimate the variance at the more relevant regions of the spectral lines. In summary, we believe the inability of \texttt{Starfish} to deliver realistic uncertainties is due to fact that their use of a Mat\'ern $\nu=3/2$ kernel is not necessarily expected to correctly describe the functional space of mismatches and the variance amplitude relevant at the location of spectral lines.



\section{Summary}
\label{sec:sum}

In this work, we have presented a new algorithm based on the spectral synthesis
technique for estimating stellar atmospheric parameters of FGK-type stars from
high resolution echelle spectra. The comparison between the data and the models
is performed  iteratively in the most sensitive zones of the spectra to changes in the atmospheric
parameters. These zones are determined after each \zaspe\ iteration and the regions
of the spectra that strongly deviate from the best model are not considered in future iterations.

\zaspe\ computes the errors and correlations in the parameters from the data itself by
assuming that the uncertainties are dominated by the systematic mismatches between
the data and the models that arise from unknown parameters of the particular atomic transitions. These systematic effects
manifest themselves by randomly modifying the strength of the absorption lines
of the synthesised spectra. The distribution of mismatches is determined by \zaspe\ from
the observed spectra and the synthetic model that produces the best fit. A random sampling method
uses an empirical distribution of line strength mismatches to modify the complete grid of synthetic spectra in a number of realisations and a new set of stellar parameters is determined in each realisation.
The complete covariance matrix can be computed from the distribution of outputs of
the random sampling method.

We have validated \zaspe\ by comparing its estimates with the SWEET-Cat catalogue of stellar parameters.
We have found that the synthetic libraries of \cite{coelho:2005} and \cite{husser:2013}
are not suitable for obtaining reliable atmospheric parameters because they present some
strong systematic trends when comparing \zaspe\ results obtained with these
grids against SWEET-Cat reference values.
We have detailed the methodology to generate our own library of synthetic spectra that we have shown is able to obtain consistent results with the SWEET-Cat catalogue. We have further confirmed the performance of our new grid by estimating the stellar parameters with \zaspe\ of three sets of
stars, whose parameters have been refined by less model dependent techniques (interferometry, asteroseismology, planetary transits).
We have estimated stellar parameters for the Sun and Arcturus using high signal-to-noise archival spectra, obtaining results consistent with state-of-the art estimates for these archetypical stars. Importantly, we obtain uncertainties that are in line with the expected level of systematic uncertainties based on studies that have performed repeat measurements of a sample of stars. Finally, we have estimated parameters for the star WASP-14, as both a way to gauge performance on a typical star that is followed-up in exoplanetary transit surveys and to compare to the \texttt{Starfish} code, the only other approach that we are aware of that deals with the systematic mismatch between models and data. Unlike \zaspe\, the \texttt{Starfish} code delivers underestimated uncertainties, a fact we believe is due to the modelling of the mismatch structure using a stationary kernel for what is fundamentally a non-stationary process as it is concentrated in the line structure.

Currently \zaspe\ works for stars of spectral type FGK. The main barriers to extend the use of \zaspe\ for stars with lower \teff\  are related to the
assumption that the systematic mismatches can be modelled by one random variable
that modifies the strength of the absorption lines. Molecular bands become the principal
feature in the spectra for stars with $\teff < 4000$ K and a more complex model
is required to characterise the systematic differences between observed and synthetic spectra. Extension to later-types will be the subject of future efforts.

\zaspe\ is mostly a \texttt{Python} based code with some routines written in \texttt{C}.
It has the option of being run in parallel with the user having the capability of entering the number of cores to
be utilised. On a 16 core CPU it takes $\approx$10 minutes for \zaspe\ to find the
synthetic spectrum that produces the best match with the data. However, to determine
the covariance matrix a couple of hours are required.
\zaspe\ has been adopted as the standard procedure for estimating the stellar atmospheric parameters
of the transiting extrasolar systems discovered by the HATSouth survey \citep{bakos:2013}; to date its results have been used for the analysis of 27 new systems \citep[from HATS-9b to HATS-35b,][]{brahm:2015,mancini:2015,ciceri:2015,brahm:2015:hs17,rabus:2016,penev:2016,deval:2016,bhatti:2016,bento:2016, espinoza:2016:hs}.
\zaspe\ is made publicly available at \url{http://github.com/rabrahm/zaspe}.

\section*{Acknowledgments}
R.B. is supported by CONICYT-PCHA/Doctorado Nacional.
R.B. acknowledges additional support from project IC120009
``Millenium Institute of Astrophysics (MAS)" of the Millennium
Science Initiative, Chilean Ministry of Economy.
 A.J. acknowledges support from FONDECYT project 1130857,
 BASAL CATA PFB-06, and project IC120009 ``Millennium Institute
 of Astrophysics (MAS)" of the Millennium Science Initiative,
Chilean Ministry of Economy.

\bibliographystyle{mnras}
\bibliography{zaspe}



\label{lastpage}

\end{document}